\documentclass[manuscript]{aastex}

\slugcomment{}

\shorttitle{Co\textbf{\emph{}}llapsed Cores in Globular Clusters}
\shortauthors{Djorgovski et al.}

\begin{document}
\title{Interstellar protons in the TeV $\gamma$-ray SNR HESS J1731-347: \\
    Possible evidence for the coexistence of hadronic and leptonic $\gamma$-rays}

\author{T.Fukuda\altaffilmark{1}, S.Yoshiike\altaffilmark{1}, H.Sano\altaffilmark{1}, K.Torii\altaffilmark{1}, H.Yamamoto\altaffilmark{1}, F.Acero\altaffilmark{2}, and Y.Fukui\altaffilmark{1}}

\affil{$^{1}$Department of Physics, Nagoya University, Furo-cho, Chikusa-ku, \\ Nagoya,
Aichi 464-8601, Japan}
\affil{$^{2}$ORAU/NASA Goddard Space Flight Center, Astrophysics Science Division, Code 661, Greenbelt, MD 20771, USA}
 
\email{tfukuda@a.phys.nagoya-u.ac.jp}

\begin{abstract}
HESS J1731-347 (G353.6-0.7) is one of the TeV $\gamma$-ray SNRs which shows the shell-like morphology. We have made a new analysis of the interstellar protons toward the SNR by using both the ${}^{12}$CO($J$=1--0) and H{\sc i} datasets. The results indicate that the TeV $\gamma$-ray shell shows significant spatial correlation with the interstellar protons at a velocity range from $-90\:\mathrm{km}\:\mathrm{s}^{-1}$ to $-75\: \mathrm{km}\:\mathrm{s}^{-1}$, and the distance corresponding to the velocity range is $\sim5.2$ kpc, a factor of 2 larger than the previous figure 3 kpc. The total mass of the interstellar protons is estimated to be $6.4\:\times10^{4}\:M_{\sun}$ , 25$\:\%$ of which is atomic gas. We have identified the cold H{\sc i} gas observed as self-absorption which shows significant correspondence with the northeastern $\gamma$-ray peak. While the good correspondence between the interstellar protons and TeV $\gamma$-rays in the north of the SNR lends support to the hadronic scenario for the TeV $\gamma$-rays, the southern part of the shell shows a break in the correspondence; in particular, the southwestern rim of the SNR shell shows a significant decrease of the interstellar protons by a factor of 2. We argue that this discrepancy can be explained as due to leptonic $\gamma$-rays, because this region well coincides with the bright shell which emit non-thermal radio continuum emission and non-thermal X-rays, suggesting that the $\gamma$-rays of HESS J1713-347 consist of both the hadronic and leptonic components. The leptonic contribution then corresponds to $\sim$20\% of the total $\gamma$-rays. The total energy of cosmic-ray protons is estimated to be $10^{49}$ erg for the $\gamma$-ray energy range of $1\:\mathrm{GeV}$-$100\:\mathrm{TeV}$ by assuming that 80\% of the total $\gamma$-ray is of the hadronic origin.

\end{abstract}


\section{Introduction}
The origin of the cosmic rays (CRs) is one of the most fundamental questions in astrophysics since the discovery of the CRs by V. Hess in 1912. 
Theoretical studies have shown that charged particles are accelerated to the energy of CRs up to $10^{15.5}$ eV via the diffusive shock acceleration (DSA) in supernova remnants (SNRs), the most promising site of acceleration in the Galaxy, where acceleration takes place efficiently between the upstream and downstream of the high velocity shock front (e.g., Bell 1978; Blandford \& Ostriker 1978). Recent observations of the TeV $\gamma$-ray SNR RX J1713.7-3946 suggest that protons may be accelerated up to an energy range of 10$\:$-$\:$800 TeV if the $\gamma$-rays are of hadronic origin (Zirakashvili \& Aharonian 2010). TeV $\gamma$-ray observations have revealed shell-like distributions of some SNRs (Aharonian et al. 2008), offering a novel opportunity to study the spatially resolved $\gamma$-rays distributions. It is an important issue to test if the $\gamma$-rays are produced by the CR protons (the hadronic process) or electrons (the leptonic process). The hadronic $\gamma$-rays are produced by the proton-proton collisions followed by neutral pion decay, while the leptonic $\gamma$-rays come from CR electrons via the inverse Compton effect. There have been considerable debates on which mechanism is mainly working in the $\gamma$-ray SNRs. In particular, keen interests focus on the four young TeV $\gamma$-ray SNRs, RX J1713.7-3946, RX J0852.0-4622, HESS J1731-347, and RCW86.

Recently, Fukui et al. (2012, 2013; hereafter F12, F13) carried out a detailed analysis of the protons in the interstellar medium (ISM) toward the young TeV $\gamma$-ray SNRs with a shell morphology, RX J1713.7-3946 and RX J0852.0-4622, and have shown that the protons in the ISM consisting of both atomic and molecular protons show a good spatial correspondence with the TeV $\gamma$-rays. These works have shown for the first time that the ISM protons have a good spatial correspondence with the TeV $\gamma$-rays, providing a necessary condition for the hadronic process, while this correspondence alone does not exclude possible contribution of a leptonic component. It is important to apply a detailed analysis of the ISM protons to the other TeV $\gamma$-ray SNRs in order to see if they show the correspondence between the ISM and the $\gamma$-ray in common.

A number of theoretical works on the $\gamma$-ray production in RX J1713.7-3946 extensively discussed the two scenarios for various physical parameters, and showed that both of the two $\gamma$-ray origins offer possible explanations on the $\gamma$-rays under different sets of physical conditions (e.g., Zirakashvili \& Aharonian 2010). One of the crucial parameters is the magnetic field; if the field is as high as 100 $\mu$G, the CR electrons suffer from strong synchrotron loss and the leptonic process may not produce the observed $\gamma$-rays (e.g., Tanaka et al. 2008). A small rapidly time-varying X-ray filament in RX J1713.7-3946 indeed suggests a mG field (Uchiyama et al. 2007). On the other hand, the $\gamma$-rays production in favor of the leptonic origin was presented by Ellison et al. (2010). These author's reasoning is that the observed $\gamma$-rays require target protons with average density more than 10 $\mathrm{cm}^{-3}$, because at density lower than this the total CR energy have to exceed that of the total kinetic energy of an SN explosion. By assuming a uniform distribution of this density, the strong shock of 3,000 km$\:\mathrm{s}^{-1}$ heats up the gas to $10^{7}$ K, which contradicts with no detection of the thermal X-rays from the hot gas in RX J1713.7-3946. The authors thus favored to exclude the hadronic origin, although they discussed also the possible role of inhomogeneous ISM for which the physical process may differ significantly. More recently, Inoue et al. (2012) have carried out magneto-hydrodynamical numerical simulations of the shock-cloud interaction by considering realistic distribution of the ISM, which is highly clumpy, consisting of dense CO clumps and inter-clump cavity (Fukui et al. 2003; Moriguchi et al. 2005; Sano et al. 2010). These authors argued that the clumpy ISM distribution is essential in understanding the origin of $\gamma$-rays and X-rays; the cavity provides an ideal site of DSA for CR acceleration and the nearby CO clumps act as the target protons which is massive enough to emit TeV $\gamma$-rays for a typical diffusion length 0.1 - 1 pc of CR protons of 10 - 800 TeV.

Since more than a decade ago, it has been discussed that a key signature for the leptonic process may be a hard $\gamma$-ray spectrum with a high-energy cut off, which shows a marked difference with that of the hadronic process having a softer spectrum with both high- and low-energy cutoffs. It has been expected that the GeV $\gamma$-ray observations can discern the two spectra. The Fermi LAT observations of RX J1713.7-3946 showed the $\gamma$-ray spectrum is hard (Abdo et al. 2011) and these authors interpreted that the spectrum lends support for the leptonic origin. It is however argued that the density-dependent penetration of CR protons is important in the clumpy CO rich environment in RX J1713.7-3946, and that the low-energy $\gamma$-rays can penetrate only into the lower density surface of the CO clumps, leading to hadronic $\gamma$-ray production with low-energy $\gamma$-ray suppressed (Zirakashvili \& Aharonian 2010; Inoue et al. 2012). Inoue et al. (2012) showed that the hadronic process for the CO distribution in RX J1713.7-3946 (Sano et al. 2010) predicts a hard $\gamma$-ray spectrum with an index of 1.5, which is equally compatible with the Fermi LAT data, and discussed that the hadronic scenario is best testable by the spatial correspondence between the $\gamma$-rays and the target ISM protons but not by the $\gamma$-ray spectrum. In addition, Inoue et al. (2012) argued that the low density inter-clump gas suppresses the thermal X-rays below the detection limit.

HESS J1731-347 was one of the unidentified TeV $\gamma$-ray sources detected by galactic plane survey with the HESS atmospheric Cerenkov telescope array (Aharonian et al. 2008). Tian et al. (2008) discovered that a new radio SNR G353.6-0.37 spatially corresponds to HESS J1731-347. The diameter of the SNR is $0\fdg 5$, allowing a comparison between the radio shell with the TeV $\gamma$-rays morphology for the HESS point spread function. The X-ray shell partly in matching with the radio shell was identified by Acero et al. (2009), Tian et al. (2010) and Bamba et al. (2012). These authors found that the X-rays are purely non-thermal, indicating that the CR electrons are accelerated up to the TeV energy range. So, HESS J1731-347 is likely a shell morphology SNR similar to RX J1713.7-3946 which is characterized by high energy TeV $\gamma$-rays and non-thermal X-rays. Given the relatively short history in studying the SNR the basic parameters of the SNR are yet not firmly established. Tian et al. (2008) suggested a distance of 3.2$\:\pm\:$0.8 kpc assuming that the H{\sc ii} region G353.42-0.37 is associated with the SNR and a similar value was confirmed as the lower limit for the distance based on the H{\sc i} absorption and CO emission (Abramowski et al. 2011). It is however to be noted that a direct link of HESS J1731-347 with ISM is still missing. It is important to identify the associated ISM to HESS J1731-347 in order to better constrain the distance and their physical parameters, and thereby to have a better understanding the $\gamma$-ray production mechanism.\\

In this paper, we present the results of an analysis of the ISM protons toward HESS J1731-347 by using both the H{\sc i} and ${}^{12}$CO($J$=1--0) data. Section 2 describes the datasets of $\gamma$-rays, 1.4 GHz continuum, H{\sc i}, and CO. Section 3 presents the results and analyses of the H{\sc i} and CO, and Section 4 gives discussion. We conclude the paper in Section 5.

\section{Datasets}
The TeV $\gamma$-rays data were obtained in the 0.1 - 100 TeV band with High Energy Stereoscopic System (H.E.S.S.). The H.E.S.S. image has a point-spread function of $0\fdg 06$ (68 \% containment radius) and the total 1 - 10 TeV $\gamma$-ray energy flux is estimated to be $6.91\times10^{-12}\:\mathrm{erg} \:\mathrm{cm}^{-2}\:\mathrm{s}^{-1}$. The present TeV $\gamma$-ray image were adopted from Abramowski et al. (2011).\\

The 21 cm H{\sc i} emission and 1.4 GHz radio continuum datasets are used from the Southern Galactic Plane Survey (SGPS) (McClure-Griffiths et al. 2005; Haverkorn et al. 2006). The SGPS images the H{\sc i} line and 1.4 GHz continuum emission with the Australia Telescope Compact Array (ATCA) and the Parkes 64 m single-dish telescope. The continuum observations have a resolution of $100^{\prime\prime}$ and a sensitivity better than 1 mJy/beam. The H{\sc i} observations have a beam size of $2\farcm2$, and the velocity resolution and typical rms noise fluctuations are $0.82\:\mathrm{km}\:\mathrm{s}^{-1}$ and 2.4 K, respectively.\\

The ${}^{12}$CO($J$=1--0) data were observed by NANTEN2 4 m telescope at Atacama in Chile in 2012 May. The observations were made by the on-the-fly mapping mode and the final pixel size of the gridded data is $60^{\prime\prime}$ with the half-power beam width (HPBW) $2\farcm7$ at the frequency of the ${}^{12}$CO($J$=1--0), 115.271 GHz. The system temperature was $\sim$ 200 K with the Superconductor-Insulator-Superconductor (SIS) receiver in the double-sideband (DSB) including the atmosphere toward the zenith. The telescope pointing was maintained within $20^{\prime\prime}$ by observing planets at 115 GHz in addition to optical observations of stars with a CCD camera equipped with the telescope. The spectrometer was a digital Fourier transform spectrometer (DFS) with 16,384 channels, providing to effective velocity resolution of $0.2\:\mathrm{km}\:\mathrm{s}^{-1}$. The final velocity resolution and typical r.m.s. roise fluctuations in the spectral analysis are 1.0 $\mathrm{km}\:\mathrm{s}^{-1}$ and 0.3 K, respectively. For the absolute intensity calibration, we observed $\rho$-Ophiuchi [$\alpha$(J2000) = $16^{\mathrm h}32^{\mathrm m}23.3^{\mathrm s}$, $\delta$(J2000) = $-24^\circ28^{\prime}39.2^{\prime\prime}$] and conversion to the main beam temperature, $T_\mathrm{mb}$ intensity was made by comparing the $T_\mathrm{mb}$ intensity observed by FCRAO (Ridge et al. 2006).

\section{Results and analyses}
\subsection{H{\sc i} and CO}
Figures 1 and 2 shows velocity channel distributions of H{\sc i} and CO overlaid with the TeV $\gamma$-rays toward HESS J1731-347. 
We show two velocity ranges where most of the ISM is distributed; one is in a range from $-90\:\mathrm{km}\:\mathrm{s}^{-1}$ to  $-70\:\mathrm{km}\:\mathrm{s}^{-1}$ and the other from $-40\:\mathrm{km}\:\mathrm{s}^{-1}$ to $0\:\mathrm{km}\:\mathrm{s}^{-1}$. The latter range includes the previous identification by Tian et al. (2008) and Abramowski et al. (2011), and the former is for the 3-kpc arm component. In a velocity range from $-80\:\mathrm{km}\:\mathrm{s}^{-1}$ to $-75\:\mathrm{km}\:\mathrm{s}^{-1}$ in Figure 1, we see an H{\sc i} hole toward the SNR, which shows a good correspondence with the outer boundary of the SNR shell. We also find the H{\sc i} is enhanced toward the SNR in a velocity range from $-90\:\mathrm{km}\:\mathrm{s}^{-1}$ to $-70\:\mathrm{km}\:\mathrm{s}^{-1}$, which may be another sign of association. A possible interpretation of the two features will be presented in Section 3.2. The CO in a velocity range from $-90\:\mathrm{km}\:\mathrm{s}^{-1}$ to $-80\:\mathrm{km}\:\mathrm{s}^{-1}$ shows two peaks at $(l, b) = (353\fdg 4, -0\fdg 5)$ and $(353\fdg 7, -0\fdg 6)$ which possibly correspond to the two $\gamma$-ray peaks in the northeast and west. Another CO feature which is extended beyond the shell boundary is seen in the east at $(l,b) = (353\fdg 8, -0\fdg 8)$. In Figure 2 the H{\sc i} distribution from $-30\:\mathrm{km}\:\mathrm{s}^{-1}$ to $0\:\mathrm{km}\:\mathrm{s}^{-1}$ shows absorption against the H{\sc ii} region G353.42-0.37 in the west of the SNR and the CO has a peak in a velocity range from $-20\:\mathrm{km}\:\mathrm{s}^{-1}$ to $10\:\mathrm{km}\:\mathrm{s}^{-1}$,  corresponding to this H{\sc ii} region. On the whole, the H{\sc i} intensity in a velocity range from $-40\:\mathrm{km}\:\mathrm{s}^{-1}$ to  $0\:\mathrm{km}\:\mathrm{s}^{-1}$ shows nearly flat distribution and the CO shows little correlated distribution with the SNR. We thus find no hint of spatial correspondence of the H{\sc i} and CO gas with the SNR in this low velocity range.

\subsection{Kinematic of the associated gas}
Figure 3 shows the H{\sc i} large-scale spatial distribution and the longitude-velocity diagram in $l=352^\circ$ -- $355^\circ$. Most of the H{\sc i} emission is in the low velocity range near $0\:\mathrm{km}\:\mathrm{s}^{-1}$ and another feature at $-90\:\mathrm{km}\:\mathrm{s}^{-1}$ to $-70\:\mathrm{km}\:\mathrm{s}^{-1}$ is the nearside 3-kpc arm, which shows a velocity gradient of $\sim4\:\mathrm{km}\:\mathrm{s}^{-1}$ per degree due to expanding and/or rotation motion. In a velocity range from $-80\:\mathrm{km}\:\mathrm{s}^{-1}$ to $-75\:\mathrm{km}\:\mathrm{s}^{-1}$ (Figure 1), the H{\sc i} surrounding the SNR shows good correspondence with the 3-kpc arm component. Figure 4 shows details of the H{\sc i} and CO longitude-velocity diagram in the 3-kpc arm in $b=-0\fdg 55$ -- $-0\fdg 48$, where a specific latitude range is chosen to show both the expanding H{\sc i} shell and CO toward the north. Toward the SNR, the H{\sc i} is seen in a velocity range from $-90\:\mathrm{km}\:\mathrm{s}^{-1}$ to $-75\:\mathrm{km}\:\mathrm{s}^{-1}$, and the CO shows a similar kinematic trend with H{\sc i}. The two CO peaks are seen toward the H{\sc i} intensity depression at $l=353\fdg 4$ and $353\fdg 7$ in Figure 4. This is natural for the lower H{\sc i} spin temperature in/near the CO gas. The CO and H{\sc i} velocities in Figure 4 tend to increase from $-85\:\mathrm{km}\:\mathrm{s}^{-1}$ to $-80\:\mathrm{km}\:\mathrm{s}^{-1}$ toward the both edges of the SNR shell. This velocity distribution is consistent with the approaching part of a shell expanding at $\sim5\:\mathrm{km}\:\mathrm{s}^{-1}$ that has only the blue-shifted side by the stellar wind acceleration by SN progenitor. The kinematics of the expanding shell is schematically shown in Appendix A. This offers a possible explanation of the enhanced H{\sc i} toward the SNR as the approaching shell on the near side in a velocity range from $-90\:\mathrm{km}\:\mathrm{s}^{-1}$ to $-80\:\mathrm{km}\:\mathrm{s}^{-1}$ (Figure 1). In this interpretation, the ISM shell is distributed only over a half of the SNR on the near side, perhaps due to the uneven initial ISM distribution, where the SN exploded on the inner edge of the nearside of the 3-kpc arm.

\subsection{Comparison with other wavelengths}
Figure 5(a) shows an overlay of the velocity averaged CO from $-90\:\mathrm{km}\:\mathrm{s}^{-1}$ to $-75\:\mathrm{km}\:\mathrm{s}^{-1}$ with the 1.4 GHz radio continuum emission. The radio emission well delineates the SNR shell with the brightest part in the south, covering a quarter of the shell. Shell-like radio emission is also seen in the northwest between the two CO peaks. The radio emission is weak toward the two CO peaks in the northeast and west. Figure 5(b) shows the velocity averaged CO and the TeV $\gamma$-rays. The two CO peaks show correspondence with the $\gamma$-rays, while the northeastern CO peak appears corresponding to only the northern half of the $\gamma$-rays peak. On the other hand, the two southern $\gamma$-ray peaks are located where the CO is not significantly detected. Figure 5(c) shows the averaged H{\sc i} and CO from $-82\:\mathrm{km}\:\mathrm{s}^{-1}$ to $-80\:\mathrm{km}\:\mathrm{s}^{-1}$ and the $\gamma$-rays. The H{\sc i} intensity appears to be enhanced toward the shell except for the northeastern $\gamma$-rays peak, where the H{\sc i} shows local depression. The H{\sc i} outside the SNR is generally extended toward the Galactic plane. Figure 6 shows more details in the northeastern H{\sc i} depression. Figure 6(a) shows an overlay of the averaged H{\sc i} intensity from $-82$ to $-80$ $\mathrm{km}\:\mathrm{s}^{-1}$ in the northeast, and Figure 6(b) and 6(c) show the averaged H{\sc i} spectrum and their distribution, respectively. We find that the H{\sc i} depression appears as flat-topped H{\sc i} profiles, and that the distribution of the H{\sc i} depression shows a good spatial correspondence with the northeastern $\gamma$-ray peak. The flat-topped profiles are in a velocity range from $-85\:\mathrm{km}\:\mathrm{s}^{-1}$ to $-78\:\mathrm{km}\:\mathrm{s}^{-1}$, which is similar to that of the CO and suggest H{\sc i} self-absorption by optically thick H{\sc i}. A quantitative analysis of the H{\sc i} depression in terms of self-absorption is presented in Section 4.\\

\section{Discussion}
\subsection{Distance of HESS J1731-347}
Distance is a basic parameter to establish physical properties of the SNR. Kinematic method based on the Galactic rotation is useful to derive a distance. The distance of HESS J1731-347 was not estimated previously by the kinematic method because the associated ISM was not identified. A recent estimate by Abramowki et al. (2011) gave only a lower limit of the distance 3.2 kpc by comparing the X-ray absorption with atomic and molecular column density. In the present work, we have shown a possible sign of spatial correspondence of the SNR with the H{\sc i} and CO in a velocity range from $-90\:\mathrm{km}\:\mathrm{s}^{-1}$ to $-75\:\mathrm{km}\:\mathrm{s}^{-1}$, and based on the assignment we infer that HESS J1731-347 is associated with the 3-kpc arm. The SN progenitor is then likely formed in the 3-kpc arm if it is a core-collapse SN. This is consistent with that a number of high-mass star forming regions have been known as maser sources in the 3-kpc arm (e.g., Green et al 2009).

Here we estimate the distance of HESS J1731-347 based on the association with the 3-kpc arm. The radius of the 3-kpc arm is estimated to be $\sim$3 kpc by assuming a circular ring based on its tangential point at $l=22^\circ$--$23^\circ$ (e.g., Woerden et al. 1957, Dame \& Thaddeus 2008). The non-circular motion of the 3-kpc arm cannot be explained by pure rotation. There are two types of interpretations for the kinematics of the 3-kpc arm; one is an expanding and rotating ring (e.g., Cohen \& Davies 1976) and the other a non-expanding elliptical structure, where the elliptical orbit is assumed to be driven by the central bar (e.g., Peters 1975). Recently, Green et al. (2011) showed that a simple elliptical ring or a circular ring with radial motion reproduces the properties of the longitude-velocity diagram of methanol masers associated 3-kpc arm better than the other published models (Mulder \& Liem 1986; Englmaier \& Gerhard 1999; Sevenster et al. 1999; Bissantz et al. 2003; Rodriguez-Fernandez \& Combes 2008). The distance of the 3-kpc arm toward HESS J1731-347 is estimated to be 6.1 kpc at $l=353\fdg 5$ by using the elliptical ring model having a rotational velocity of $320\:\mathrm{km}\:\mathrm{s}^{-1}$ with a semi-major axis of 4.1 kpc and a semi-minor axis of 2.2 kpc at an orientation of the major axis $38^\circ$ to the line of sight in the positive Galactic-longitude. On the other hand, the distance of the 3-kpc arm is estimated to be 5.2 kpc in the expanding circular ring model with a radius of 3.4 kpc from the Galactic centre, radial velocity of $48\:\mathrm{km}\:\mathrm{s}^{-1}$, and rotational velocity of $201\:\mathrm{km}\:\mathrm{s}^{-1}$. So, we estimate that the distance of HESS J1731-347 is in a range of $5.2$--$6.1$ kpc. This result is consistent with the distance of a methanol maser 9.621+0.196 in the 3-kpc arm, $5.2\pm0.6$ kpc determined by astrometric parallax (Sanna et al. 2009) and the lower limit derived from the previous works (Tian et al. 2008; Abramowski et al. 2011). We shall hereafter apply a distance of 5.2 kpc to calculate lower limits of the physical parameters of the SNR.

By adopting the distance, the radius of the SNR is $\sim22$ pc ($r=0\fdg 24$) and the age and the shock speed are estimated to be $\sim$4,000 yr and $\sim$2,100 $\mathrm{km}\:\mathrm{s}^{-1}$ using Sedov-Taylor solution for an assumed SN explosion of $10^{51}$ erg and a density of 0.01 $\mathrm{cm}^{-3}$ as the upper limit of density to the non-thermal X-ray feature in case of electron plasma temperature $kT_e = 1\:\mathrm{keV}$ (Abramowski et al. 2011). On the other hand, the TeV luminosity of HESS J1731-347 in the $1$--$30$ TeV energy band is estimated to be $2.8\times(d/5.2\:\mathrm{kpc})^{2}\times10^{34}\ \mathrm{erg}\ \mathrm{s}^{-1}$ from the published $\gamma$-ray data (Abramowski et al. 2011). This luminosity is approximately three times as large as that of RX J1713.7-3946. HESS J1731-347 is therefore one of the most powerful TeV $\gamma$-ray sources known to date.

\subsection{The ISM proton distribution}
We shall here estimate the total ISM protons in the SNR shell. The total intensity of ${}^{12}$CO($J$=1--0), $\mathrm{W}({}^{12}\mathrm{CO})$, can be converted into the molecular column density $N_\mathrm{H_2}\:[\mathrm{cm}^{-2}]$. The molecular column density is estimated by $N_\mathrm{H_2} = X_\mathrm{CO} \times \mathrm{W}({}^{12}\mathrm{CO})$, where the $X_\mathrm{CO}$ factor is adopted as $X_\mathrm{CO} = 2.0\:\times\:10^{20}\:[\mathrm{cm}^{-2}\:\mathrm{{K}^{-1}\:{km}^{-1}}\:\mathrm{s}]$ (Bertsch et al. 1993). Then, the proton column density in molecular form is given as $N_\mathrm{p}(\mathrm{H_2}) = 2 \times N_\mathrm{H_2}$. The typical H{\sc i} peak intensity is around 50 K toward the SNR and the atomic proton column density is estimated by assuming that the H{\sc i} is optically thin. The H{\sc i} column density $N_\mathrm{p}$(H{\sc i}) [$\mathrm{cm}^{-2}$] is given as $N_\mathrm{p}(\mbox{H{\sc i}}) = 1.823 \times 10^{18} \int\mathrm{T_b}\:dV$ (Dickey \& Lockman 1990), where $T_b$ [K] and $V$ [$\mathrm{km}\:\mathrm{s}^{-1}$] are the H{\sc i} intensity and the velocity. From the present results the velocity range of The H{\sc i} and ${}^{12}\mathrm{CO}$ is taken to be from $-90\:\mathrm{km}\:\mathrm{s}^{-1}$ to $-75\:\mathrm{km}\:\mathrm{s}^{-1}$. Then, the total ISM proton column density is given as $N_\mathrm{p}(\mathrm{H_2}+\mbox{H{\sc i}})  = N_\mathrm{p}(\mathrm{H_2}) + N_\mathrm{p}(\mbox{H{\sc i}})$. The averaged proton column density in whole shell derived for optically thin H{\sc i} is $1.1\:\times\:10^{21}\:\mathrm{cm}^{-2}$, while that of molecular gas is $4.3\:\times\:10^{21}\:\mathrm{cm}^{-2}$. The total proton mass are $1.3\:\times10^{4}\:M_{\sun}$ and $5.1\:\times10^{4}\:M_{\sun}$ for atomic gas and molecular gas, respectively, in the entire SNR defined by a radius of $0\fdg 24$ (Figure 7). This indicates that the atomic proton corresponds to about $25\:\%$ of the total ISM protons, and that the ISM protons associated with the SNR are dominated by $\mathrm{H_2}$.

We note that the H{\sc i} emission may not be optically thin in some part of the H{\sc i} gas. We find that the HI shows intensity depression toward the south of the northeastern CO peak, which may suggest self-absorption by cold H{\sc i} gas(Figure 6(a)). If the spin temperature of H{\sc i} is lower than $\sim50$ K, the cold H{\sc i} can cause the self-absorption. The cold H{\sc i} is probably in lower density gas which cannot emit CO emission.  The self-absorption dips vary significantly from point to point in Figure 6(c), indicating that the H{\sc i} optical depth may not be very large. In Figure 6(c) the H{\sc i} self-absorption is seen from $-85\:\mathrm{km}\:\mathrm{s}^{-1}$ to $-79\:\mathrm{km}\:\mathrm{s}^{-1}$, while the CO velocity from $-88\:\mathrm{km}\:\mathrm{s}^{-1}$ to $-73\:\mathrm{km}\:\mathrm{s}^{-1}$  peaked at $-80\:\mathrm{km}\:\mathrm{s}^{-1}$ is somewhat red-shifted by a few $\mathrm{km}\:\mathrm{s}^{-1}$ from the H{\sc i} self-absorption. The velocity and shape of the H{\sc i} self-absorption are not well matched with the CO. This suggests that the CO is not in the foreground of the H{\sc i} but is inside or on the far-side of the H{\sc i}, although the CO and cold H{\sc i} are probably physically in close contact with each other as inferred from the spatial distribution(Figure 6(c)). The $\gamma$-ray contour is extended toward the H{\sc i} depression, showing possible correspondence between the $\gamma$-rays and the cold H{\sc i}. The H{\sc i} intensity $T_L(V)$ is expressed as follows (e.g., Sato \& Fukui 1978):

\begin{equation}
T_L(V)=T_s[1-e^{-\tau(V)}]+T_{L}^{\mathrm{FG}}(V)+[T_{L}^{\mathrm{BG}}(V)+T_{C}^{\mathrm{BG}}]e^{-\tau(V)}-(T_{C}^{\mathrm{FG}}+T_{C}^{\mathrm{BG}})
\end{equation}

where $T_L(V)$, $T_s$, $\tau(V)$, ${T_{L}^{\mathrm{FG}}}(V)$, and ${T_{L}^{\mathrm{BG}}}(V)$ are the observed H{\sc i} intensity, the spin temperature, the optical depth of cold H{\sc i} in the cloud, and the foreground and background H{\sc i} brightness temperature, respectively, at velocity $V$. $T_{L}^{\mathrm{FG}}(V)$ and ${T_{L}^{\mathrm{BG}}}(V)$ are the continuum brightness temperature at 1.4 GHz in the foreground and background of the cloud, respectively. The radio continuum emission is weak toward the northeastern $\gamma$-ray peak, and ${T_{C}^{\mathrm{FG}}}(V)$ and ${T_{C}^{\mathrm{BG}}}(V)$ are nearly zero compared with $T_L(V)$. We shall attempt to roughly estimate the cold H{\sc i} column density toward the northeast $\gamma$-ray peak by using the H{\sc i} spectrum in Figure 6(c). The background H{\sc i} intensity ${T_{L}^{\mathrm{BG}}}(V)$ is given by the averaged spectra around the absorption area. The spin temperature $T_s$ of the cold H{\sc i} gas is estimated $T_s$ to be less than 34 K from the lowest H{\sc i} brightness at the $-80$ $\mathrm{km}\:\mathrm{s}^{-1}$ in Figure 6(c) and probably higher than 20 K which corresponds to $T_s$ at density $100\:\mathrm{cm}^{-3}$(e.g., Goldsmith et al. 2007).  These two cases for $T_s$ are consistent with the results that the optically thick H{\sc i} is characterized by $T_s$ in a range from 15 K to 40 K for 77 \% of H{\sc i} in the whole sky at $|b|>15^\circ$ (Fukui et al. 2014a and Fukui et al. 2014b). Then, the cold H{\sc i} column density is given as $N_\mathrm{p}(\mbox{H{\sc i}}) = 1.823 \times 10^{18} T_s \int\tau(V)\:dV$. 

We calculate for two extreme cases the cold H{\sc i} column density to be $N_\mathrm{p}(\mbox{H{\sc i}}) = 1.4 \times 10^{20}\:\mathrm{cm}^{-2}$($\tau = 0.5$) and $N_\mathrm{p}(\mbox{H{\sc i}}) = 1.6 \times 10^{21}\:\mathrm{cm}^{-2}$($\tau = 4.0$) for assumed $T_s$ = 20 K and 34 K, respectively. The cold H{\sc i} column density corresponds to $15$ \% -- $180$ \% of the H{\sc i} column density derived for the optically thin H{\sc i}. The cold H{\sc i} mass is estimated to be $70$--$800\:M_{\sun}$ within the 6$\sigma$ contour of the $\gamma$ rays(Figure 6(c)), which correspond to 2 \% -- 25 \% of the CO mass, and suggest that the cold H{\sc i} can locally affect the distribution of the hadronic $\gamma$-rays if the H{\sc i} optically depth is $\gtrsim1$ .

\subsection{Comparison between the ISM protons and the $\gamma$-rays}
Figure 7(a) presents the sum of the molecular and atomic protons, the total ISM protons superposed on the TeV $\gamma$-rays, and Figure 7(b) an overlay of the total ISM protons and the radio continuum emission. Spatial correspondence between the $\gamma$-rays and ISM protons is key to discern the $\gamma$-ray production mechanism according to Inoue et al. (2012), F12 and F13. The two CO peaks with the cold H{\sc i} show relatively good correspondence with the two $\gamma$-ray peaks, and it is a reasonable assumption that the $\gamma$-rays are created by the hadronic process there. We show the normalized azimuthal proton column density and the $\gamma$-ray surface brightness in Figure 8(a), where the position angle $0^\circ$ corresponds to the north and $90^\circ$ to the east in the equatorial coordinate (Figure 7(a)). The TeV $\gamma$-rays are integrated within a radius $0\fdg 3$ centered on the SNR, $(l, b) = (353\fdg 54, -0\fdg 67)$ following Abramowski et al .(2011). The $\gamma$-rays are peaked at a radius of $0\fdg 24$ and the ISM protons are integrated within this peak radius following F12 and F13 in order to include interacting protons responsible for in the $\gamma$-rays production. The error bars in the $\gamma$-ray profiles are adopted by Abramowski et al. (2011), and those in protons are 10\% of the average proton column density mainly due to the calibration uncertainty in CO. 

In the azimuthal distribution an eye inspection indicates that six points in $-126^\circ$ -- $+54^\circ$ toward the two CO peaks show a relatively good match with the $\gamma$-rays. By assuming that the $\gamma$-rays are produced predominantly by the hadronic process, we made a least-squares fitting to minimize the deviation between the ISM protons and $\gamma$-rays for the six points. This gives tentatively the ratio between the $\gamma$-rays surface brightness(arbitrary unit) and the ISM protons column density as 6.5. Although the $\gamma$-rays error bars are not small, the profile in the north shows that the ISM protons correspond reasonably well to the TeV $\gamma$-rays within the error range. The ISM protons in the remaining four points with yellow color in Figure 8(a) show a factor of $\sim2$ deficiency relative to the $\gamma$-rays. This deficiency indicates that the spatial correspondence between the ISM protons and the $\gamma$-rays is not complete over the whole shell. The same ratio is assumed in Figure 8(b), a comparison between the $\gamma$-rays and ISM in the radial distribution. The overall correspondence between the interstellar protons and the $\gamma$-rays seems fairly good in the radial distribution within the shell. We note that the radial distribution shows a small discrepancy between the $\gamma$-rays and the total ISM protons due to the deficiency of the ISM protons in the south (Figure 8(a)). In summary, we see generally good spatial correspondence between the $\gamma$-rays and the ISM protons as expected in case of the hadronic $\gamma$-rays. If we assume that about 50 \% of the $\gamma$-rays in the four southern points, which corresponds to the ISM protons in Figure 8(a), are due to the hadronic process, it is possible that about 80 \% of the total $\gamma$-rays are ascribed to the hadronic origin. We will discuss later that the remaining 20\% of the $\gamma$-rays are possibly ascribed to the leptonic process. 

We discuss further if the density-dependent penetration of CR protons may affect the above argument. The typical penetration length of CR protons is given as follows (Inoue et al. 2012).

\begin{equation}
l_\mathrm{pd}\ \sim\ 0.1\ \eta^{1/2}\ \left(\frac{E}{10\:\mathrm{TeV}}\right)^{1/2}\ \left(\frac{B}{100\:\mu \mathrm{G}}\right)^{-1/2}\ \left(\frac{t_\mathrm{age}}{10^{3}\:\mathrm{yr}}\right)^{1/2}\ [\mathrm{pc}]. 
\end{equation}

where $E$, $B$, and $t_\mathrm{age}$ are the particle energy, the magnetic field and the age of the SNR. Another parameter $\eta$ is given by ${B^{2}}/{\delta B^{2}}$ the degree of magnetic field fluctuations. In case of RX J1713.7-3946, $\eta\sim1$ is reasonable at least around the CO clumps (e.g., Uchiyama et al. 2007). For an energy of 10 TeV, magnetic field of 10 $\mu$G, and age of 4,000 yr, the length is estimated to be 0.6 pc, corresponding to $\sim30^{\prime\prime}$ at 5.2 kpc. The present CO data have a spatial resolution of $\sim4$ pc, too coarse to see such penetration depths, whereas it is generally the case that the CO clouds have clumpy distribution less than sub-pc (e.g., Sano et al. 2010). Recent CO observations with Mopra by us show such small-scale clumpy distribution of CO at $30^{\prime\prime}$ resolution (unpublished). We therefore suggest that the CR protons are mostly interacting with the ISM protons, although the densest molecular gas may not be reached by lower-energy CR protons as argued by Zirakashvili \& Aharonian (2010) and Inoue et al. (2013).

The ISM protons are dominated by the molecular form in HESSJ1731-347 and the average proton density both molecular and atomic over the whole SNR is estimated to be $60\:\mathrm{cm}^{-3}$. For the points where the $\gamma$-rays match the ISM protons the average density is $80\:\mathrm{cm}^{-3}$, while for the other four points the average density is $40\:\mathrm{cm}^{-3}$. The density where the $\gamma$-rays match the ISM protons is similar to the average density of target protons $100\:\mathrm{cm}^{-3}$ found in RX J1713.7-3946 and RX J0852.0-4622 (F12 and F13). The present study therefore offers possibly a third case which supports the hadronic origin of the $\gamma$-rays by the spatial correspondence. Assuming that the molecular cloud is being fully penetrated by CR protons, the total energy of CR protons ($E\:\geq1\:\mathrm{GeV}$) for a spectral slope of 2.0 is estimated to be $\mathrm{W_p}=7\times10^{48}\:(n/60\:{\mathrm{ cm^{-3}})}^{-1}{(d/5.2\:\mathrm{kpc})}^2$ erg by adopting 80 \% of the total $\gamma$-ray flux from sub-region of the SNR (Abramowski et al. 2011), corresponding to 1\% of the total kinetic energy of a SNe $\sim10^{51}$ erg. 

A comparison between RX J0852.0-4622, RX J1713.7-3946 and HESS J1731-347 is given in Table 1. In RX J1713.7-3946 the molecular and atomic masses are $10^{4}\:M_{\sun}$ for each. In RX J0852.0-4622, the atomic mass is $10^{4}\:M_{\sun}$, while the molecular mass is $10^{3}\:M_{\sun}$(F13). The atomic proton mass are not much different among three SNRs, but the molecular proton mass is significantly different. In particular, HESS J1731-347 have a large mass of $5.1\times\:10^{4}\:M_{\sun}$ as target for CR protons, which is consistent with the large TeV $\gamma$-ray luminosity if the $\gamma$-rays are produced predominantly by the hadronic process. We find that these three young TeV $\gamma$-ray SNRs have the total CR protons in the order of $10^{48}$-$10^{49}$ erg. The energy suggests that the CR acceleration efficiency is 0.1-1 \% in such a young stage.

We suggest that the apparent small energy of CR protons may be in part due to the inhomogeneous ISM distribution which hampers a full coupling of the CRs with the ISM. The escaping of the CRs may also be significant and has not been measured yet in the three SNRs (see for W44, Uchiyama et al. 2013). In addition, the efficiency of CR acceleration may possibly grow in time and this can be tested by comparing those with the middle-aged SNRs having an age of $\sim10^{4}$ yr (e.g., Giuliani et al. 2010; Yoshiike et al. 2013). By considering all these effects, the apparent efficiency of CR acceleration $0.1$--$1$ \% may be considered as secure lower limits for the actual efficiency, and we suggest that the net efficiency can be significantly larger than $0.1$--$1$ \% over the whole lifetime of the SNRs.\\

\subsection{A leptonic component in the south shell}
Figures 7 and 8 indicate that the major part of the shell in HESS J1731-347 is@consistent with the hadronic origin since the spatial correspondence between the hadronic $\gamma$-rays and the ISM protons is naturally expected (Inoue et al. 2012). We find however a significant deviation by a factor of 2 between them in nearly half of the shell in the south. The weak or no CO emission poses an upper limit for the column density, and the low ISM column density toward the southern part of the shell in Figures 7 and 8 is robust.

Figure 7(b) indicates that the southern shell is toward the significant enhancement of the radio continuum synchrotron emission. This invokes a possibility that the leptonic process is responsible for the gap between the ISM protons and the $\gamma$-rays. The region has significantly lower H{\sc i} density $\sim10\ \mathrm{cm}^{-3}$ than the rest of the SNR and the interaction between the shock waves and the ISM is weaker than elsewhere. This leads to weaker magnetic fields and lower target ISM protons in the shock-cloud interaction (Inoue et al. 2012). The weaker magnetic field causes weaker synchrotron cooling and favors the leptonic $\gamma$-rays. If this is the case, we set up a picture that the $\gamma$-rays in HESS J1731-347 consist of mainly the hadonic $\gamma$-rays over the whole SNR, but the contribution of the leptonic $\gamma$-rays is significant, at a level similar to the hadronic $\gamma$-rays in the southern shell. Given the significant difference by a factor of $\sim$2 between the ISM protons and $\gamma$-rays, such a leptonic component may be responsible for $\sim20$ \% of the total $\gamma$-ray flux of HESS J1731-347. The Suzaku X-ray distribution lends support for this explanation, because the southern shell, where the radio non-thermal emission is strong, also emits strong non-thermal X-rays at 2 - 10 keV (Bamba et al. 2012). Considering that the CR electrons emitting the 2 - 10 keV X-rays have energy in the TeV range, we find it reasonable that such electrons are also responsible for the TeV $\gamma$-rays via the inverse Compton effect.

\section{Conclusions}
We summarize the present paper as follows.
\begin{description}
\item[(1)]A comparison between H{\sc i}, ${}^{12}$CO($J$=1--0), TeV $\gamma$-rays and 1.4 GHz radio continuum emission has shown that the SNR shell of HESS J1731-347 is associated with the CO and H{\sc i} in a velocity range from $-90\:\mathrm{km}\:\mathrm{s}^{-1}$ to $-75\:\mathrm{km}\:\mathrm{s}^{-1}$. The H{\sc i} shows a hole-like depression at $-85\:\mathrm{km}\:\mathrm{s}^{-1}$ as well as enhanced emission at $-80\:\mathrm{km}\:\mathrm{s}^{-1}$ toward the shell, suggesting that the SNR shell is associated with the H{\sc i}. We suggest that the H{\sc i} is possibly on the nearside of the shell and is expanding at $\sim5\:\mathrm{km}\:\mathrm{s}^{-1}$. The CO well corresponds to the northern part of the shell but has no obvious counter part in the southern part. The velocity from $-90\:\mathrm{km}\:\mathrm{s}^{-1}$ to $-75\:\mathrm{km}\:\mathrm{s}^{-1}$ corresponds to the nearside of the 3-kpc expanding arm. The distance of the SNR is estimated to be in a range of 5.2 - 6.1 kpc compatible with the lower limit derived from Tian et al. (2008) and Abramowski et al. (2011), if the association with the 3-kpc expanding arm is adopted. Then the lower limit of radius, age, and TeV luminosity of the SNR are estimated to be $\sim$22 pc, $\sim$4,000 yr, and $2.8\times(d/5.2\  \mathrm{kpc})^{2}\times10^{34}\ \mathrm{erg}\ \mathrm{s}^{-1}$ in the 1-30 TeV energy band, respectively. It should be noted that this luminosity is approximately three times as large as that of RX J1713.7-3946. 

\item[(2)]The H{\sc i} and CO data are combined as the total ISM protons and the radial and azimuthal profiles are derived. Although the $\gamma$-rays error bars are large,  these profiles show generally that the ISM protons correspond to the TeV $\gamma$-rays. The $\gamma$-rays show a flat azimuthal distribution in the whole shell, whereas the ISM protons in the southern shell show a factor of $\sim2$ deficiency. 

\item[(3)]The spatial correspondence between the ISM protons and $\gamma$-rays lends support for the hadronic origin of the $\gamma$-rays. The average ISM density in HESS J1731-347 where the correspondence is good is $\sim100\ \mathrm{cm}^{-3}$. The total CR proton energy in the hadronic scenario is then derived to be $\sim10^{49}$ erg, a factor of a few greater than that of the two TeV $\gamma$-rays SNRs, RX J1713.7-3946 and RX J0852.0-4622. 

\item[(4)]The southern shell shows a significant decrease in the ISM protons and the average density H{\sc i} is as low as $10\ \mathrm{cm}^{-3}$. We suggest that this low density of the interacting gas may lead to lower amplification of the magnetic fields and thereby favors the leptonic origin of the $\gamma$-rays. This is consistent with the enhanced non-thermal X-rays and radio emission toward the southern shell. If this is the case, the $\gamma$-rays of HESS J1713-347 consist of both the hadronic and leptonic components. The leptonic contribution then corresponds to 20\% of the total $\gamma$-rays. 
\end{description}

\acknowledgments
NANTEN2 is an international collaboration of 10 universities: Nagoya University, Osaka Prefecture University, University of Cologne, University of Bonn, Seoul National University, University of Chile, University of New SouthWales, Macquarie University, University of Sydney, and University of ETH Zurich. This work is financially supported by a grantin-aid for Scientific Research (KAKENHI, No. 15071203, No.21253003, No. 20244014, No. 23403001, No. 22540250,
No.22244014, No. 23740149-01, and No. 22740119) from MEXT(the Ministry of Education, Culture, Sports, Science and Technology of Japan). This work is also financially supported by the Young Research Overseas Visits Program for Vitalizing Brain Circulation (R2211) and the Institutional Program for Young Researcher Overseas Visits (R29) by JSPS (Japan Society for the Promotion of Science) as well as the JSPS core-to-core program (No. 17004). We also
acknowledge the support of the Mitsubishi Foundation and the Sumitomo Foundation. This research was supported by the grant-in-aid for Nagoya University Global COE Program, gQuest for Fundamental Principles in the Universe: From Particles to the Solar System and the Cosmos from MEXT. The satellite internet connection for NANTEN2 was provided by the Australian Research Council.
\clearpage

\appendix
\section{Expanding motion of the ISM and interpretation of position-velocity diagram}
In Figure 9(a,b), we show an expanding spherical shell model of radius $R_{0}\sim22\mathrm{pc}$ and uniform expansion velocity $V_{0} \sim5\:\mathrm{km}\:\mathrm{s}^{-1}$ and its position-velosity diagram. If the expanding shell uniformly spread to space, we can expect to observe both the red-shifted and blue-shifted expanding motion. In the case of HESS J1731-347, we observed shell expanding motion that has only the blue-shifted side. This may be due to the low density ISM on the farside of the expanding shell. Observed H{\sc i} expanding velocity $\sim$5 $\mathrm{km}\:\mathrm{s}^{-1}$ for mass of $1\:\times\:10^{4}\:\:M_{\sun}$ corresponds the kinetic energy of $\sim 1\times\:10^{48}$ erg. This is consistent with the stellar wind acceleration by the SN progenitor over 1 Myr (see e.g., Fukui et al. 2012). The expansion of the ISM shell is described as follows (eg. Caster et al.1975; Weaver et al. 1977). The radius and velocity of the cavity are determined by the initial phase when the radiation loss is not effective as given below;

\begin{equation}
\mathrm{R}(t)=27 \times \left(\frac{n}{1\:\mathrm{cm}^{-3}}\right)^{-1/5} \left(\frac{L_{w}}{10^{6}\:\mathrm{erg\:s^{-1}}}\right)^{1/5} \left(\frac{t_{w}}{10^{6}\:\mathrm{yr}}\right)^{3/5} [\mathrm{pc}] 
\end{equation}
\begin{equation}
V(t)=16 \times \left(\frac{n}{1\:\mathrm{cm}^{-3}}\right)^{-1/5} \left(\frac{L_{w}}{10^{6}\:\mathrm{erg\:s^{-1}}}\right)^{1/5} \left(\frac{t_{w}}{10^{6}\:\mathrm{yr}}\right)^{-2/5} [\mathrm{km}\:\mathrm{s}^{-1}] 
\end{equation}

where $n$ is the gas desnity, $L_{w}$ is stellar wind mechanical luminosity, and $t$ is wind lifetime. For $n =20\:\mathrm{cm}^{-3}$, $L_{w}=10^{6}\:\mathrm{erg s^{-1}}$, $t_{w}=2\times10^{6}\:\mathrm{yr}$, the radius is estimated $\sim$22 pc, with expanding velocity $\sim$5 $\mathrm{km}\:\mathrm{s}^{-1}$.\\

\clearpage

\begin{table}
\begin{center}
\caption{A Comparison between RX J0852.0-4622, RX J1713.7-3946 and HESS J1731-347}
\scalebox{0.8} [0.8]{
\begin{tabular}[t]{ccccccccc}
\tableline\tableline
$\:$ & $\mathrm{RX J0852.0-4622^a}$ & $\mathrm{RX J1713.7-3946^b}$ & $\mathrm{HESS J1731-347^c}$ \\

\tableline
Distance$\:$(kpc) & 0.7 & 1 & $\mathrm{5.2^d}$ \\
Radius$\:$(pc) & 13 & 9 & 22 \\
Age$\:$(years) & 1700 & 1600 & 4000 \\
Atomic proton mass$\:$($10^{4}\:\:M_{\sun}$) & 1 & 1 & 1.3 \\
Molecular proton mass$\:$($10^{4}\:\:M_{\sun}$) & 0.1 & 1 & 5.1 \\
Total proton mass$\:$($10^{4}\:\:M_{\sun}$) & 1.1 & 2 & 6.4 \\
Average density$\:$($\rm{cm^{-3}}$) & 40 & 100 & 60 \\
$L_{\gamma}\:(1-10\mathrm{TeV})$ $\:$($10^{34}\:\:\rm{erg\:\:s^{-1}}$) & 0.63 & 0.81 & 2.8 \\
Total CR proton energy (erg) & $\sim10^{48}$ & $\sim10^{48}$ & $\sim10^{49}$ \\ 
\tableline
\end{tabular}
}
\end{center}
\footnotetext{}{\bfseries{Notes.}}\\
\footnotetext{}{$^{\rm{a}}$ Fukui et al.(2013)}\\
\footnotetext{}{$^{\rm{b}}$ Fukui et al.(2012)}\\
\footnotetext{}{$^{\rm{c}}$ Present work}\\
\footnotetext{}{$^{\rm{d}}$ We estimate that the distance is in a range of $5.2$ kpc -- $6.1$ kpc. Here we apply a distance of 5.2 kpc to calculate lower limits of the physical parameters of the SNR.}\\

\end{table}
\clearpage

\begin{figure}
\plotone{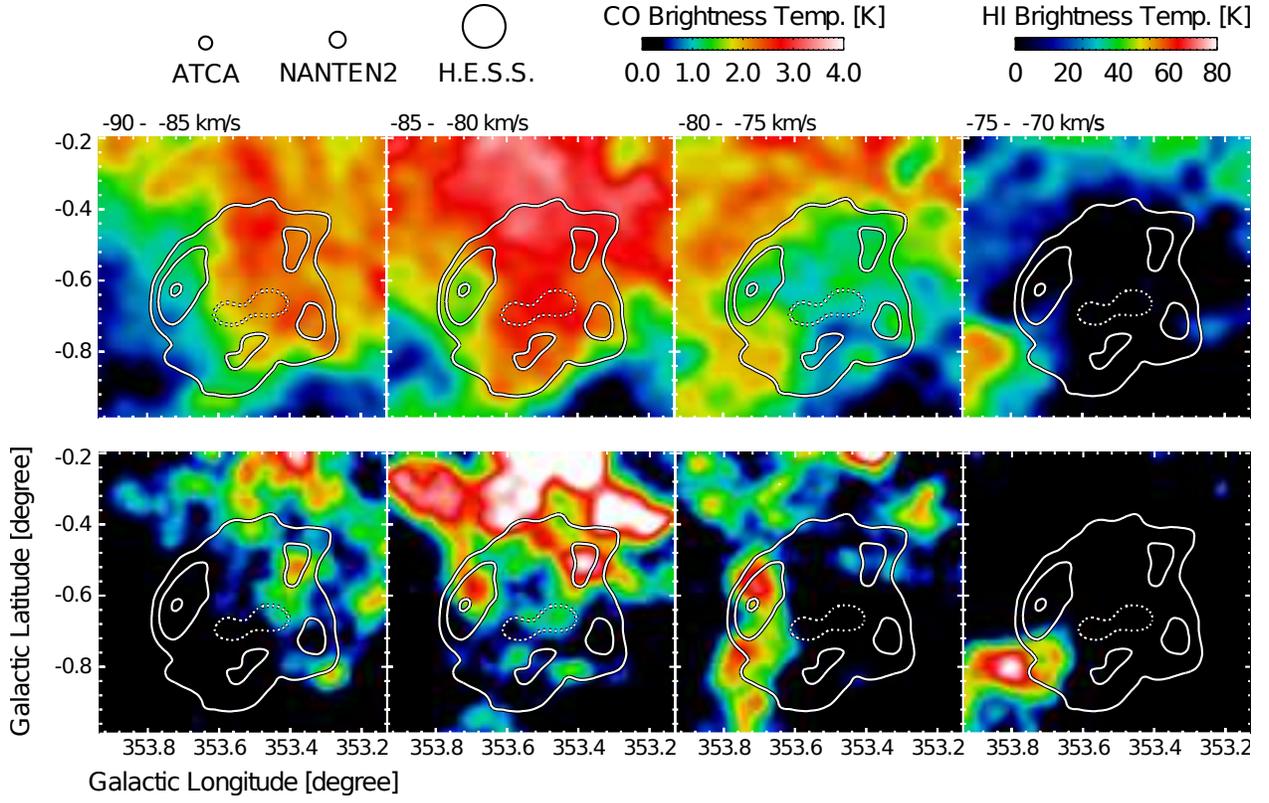}
\caption{Velocity channel map of H{\sc i} and ${}^{12}$CO($J$=1--0) overlaid on the TeV $\gamma$-ray distribution. Top: H{\sc i} image superposed on the TeV $\gamma$-ray contours. TeV $\gamma$-rays contour levels are 4, 6, 8$\sigma$. Bottom: ${}^{12}$CO($J$=1--0) image superposed on the TeV $\gamma$-ray contours. Each panel show H{\sc i} and  ${}^{12}$CO($J$=1--0) distribution every 5 $\mathrm{km}\:\mathrm{s}^{-1}$ in a velocity range from -90 to -70 $\mathrm{km}\:\mathrm{s}^{-1}$.} 
\end{figure}
\clearpage

\begin{figure}
\plotone{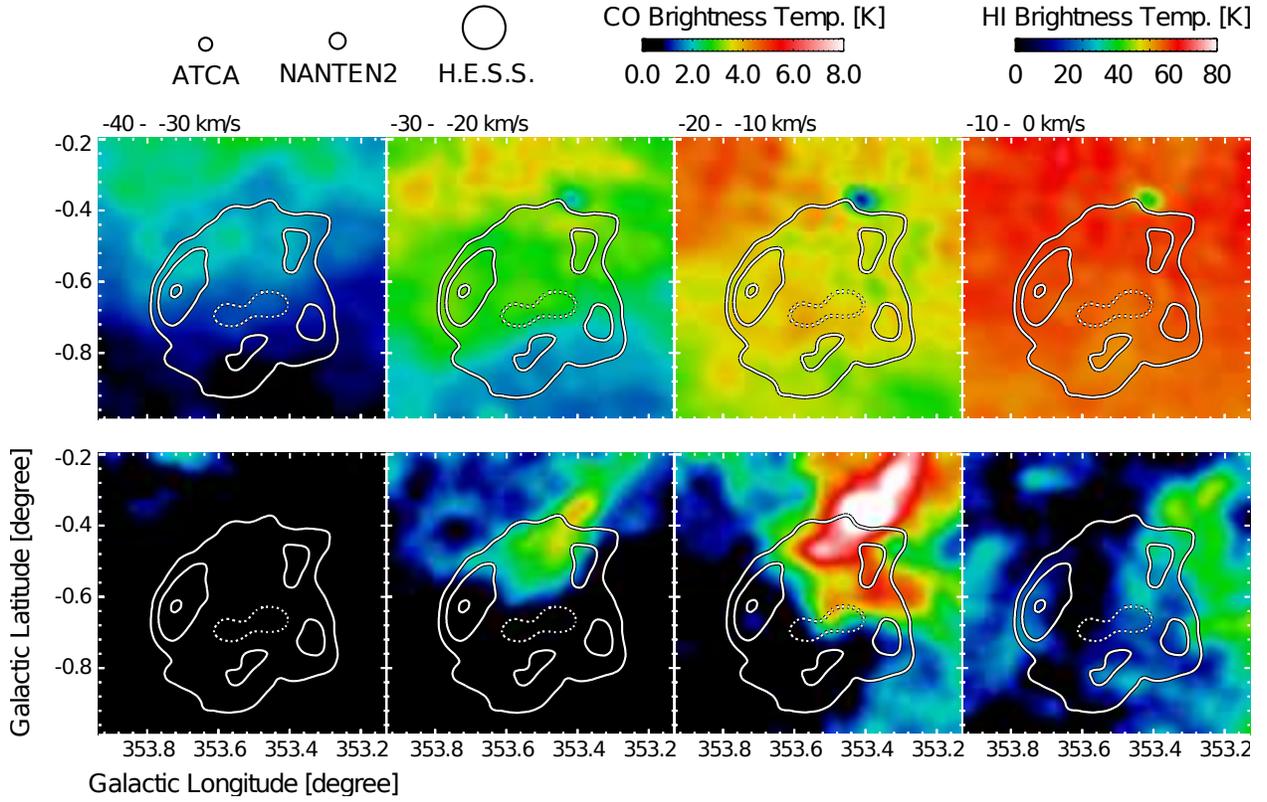}
\caption{Velocity channel map of H{\sc i} and ${}^{12}$CO($J$=1--0) overlaid on the TeV $\gamma$-ray distribution in common with Figure 2. Each panel show ${}^{12}$CO($J$=1--0) and H{\sc i} distribution every 10 $\mathrm{km}\:\mathrm{s}^{-1}$ in a velocity range from $-40$ to $0\:\mathrm{km}\:\mathrm{s}^{-1}$.} 
\end{figure}
\clearpage

\begin{figure}
\begin{center}
\includegraphics[scale=1.0,angle=0,clip]{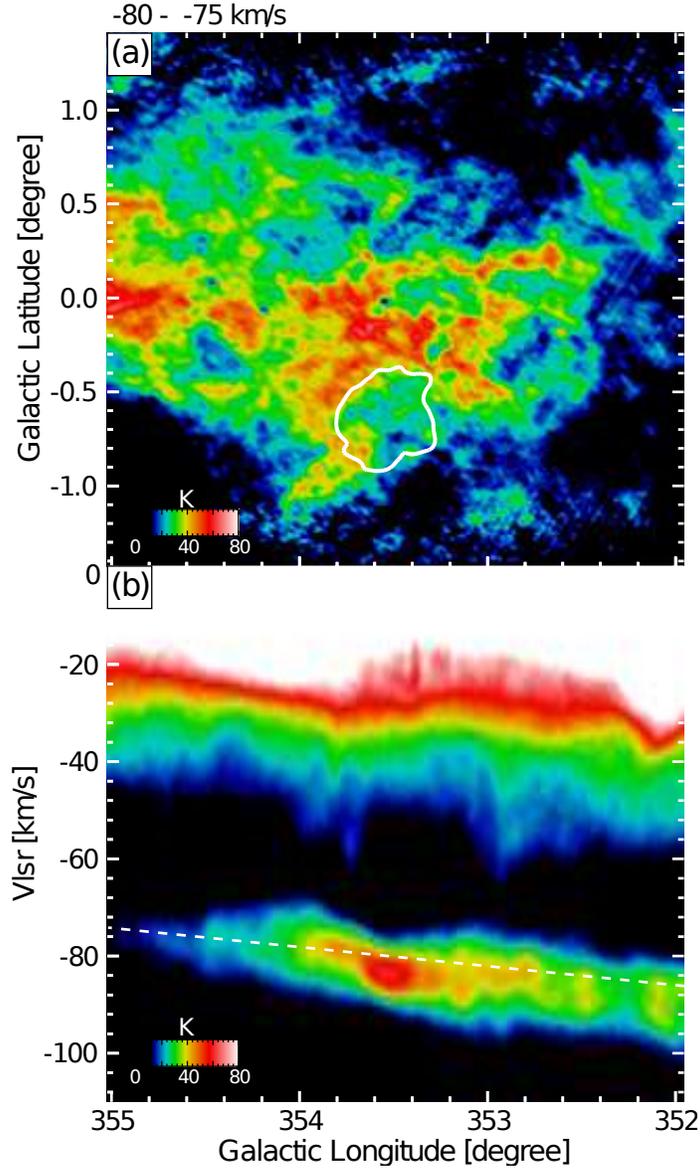}
\caption{Large scale distribution of H{\sc i}. (a)Averaged intensity distribution of H{\sc i} in a velocity range from $-80\:\mathrm{km}\:\mathrm{s}^{-1}$ to $-75\:\mathrm{km}\:\mathrm{s}^{-1}$ is shown in color. The $\gamma$-ray boundary of HESS J1731-347 is indicated by white line. (b)Galactic longitude -- velocity map of H{\sc i} in the same region as Figure 4(a). The integrated range in Galactic latitude is from $-0\fdg 93$ to $-0\fdg 37$ and is the same as diameter of $\gamma$-ray shell. The position of the 3-kpc expanding arm is indicated by dashed white lines applied as $v\:[\mathrm{km}\:\mathrm{s}^{-1}] = -53.1 + 4.16\:l[\mathrm{degree}]$ (Dame \& Thaddeus 2008). } 
\end{center}
\end{figure}
\clearpage

\begin{figure}
\begin{center}
\includegraphics[scale=1.2,angle=0,clip]{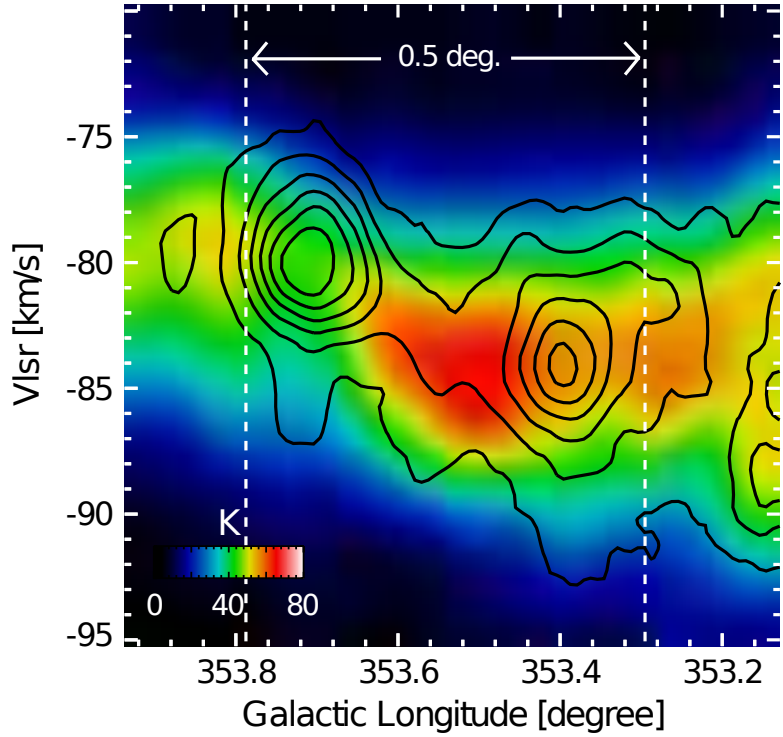}
\caption{The detail of longitude -- velocity map of H{\sc i}(Color) and ${}^{12}$CO($J$=1--0)(contour). The integrated range in Galactic latitude is from $-0\fdg 55$ to $-0\fdg 48$. The minimum contour levels and intervals of ${}^{12}$CO($J$=1--0) is 0.4 K and 0.6 K, respectively. The $\gamma$-ray shell diameter of HESS J1731-347 is indicated by dashed line.} 
\end{center}
\end{figure}

\clearpage

\begin{figure}
\begin{center}
\includegraphics[scale=0.4,angle=0,clip]{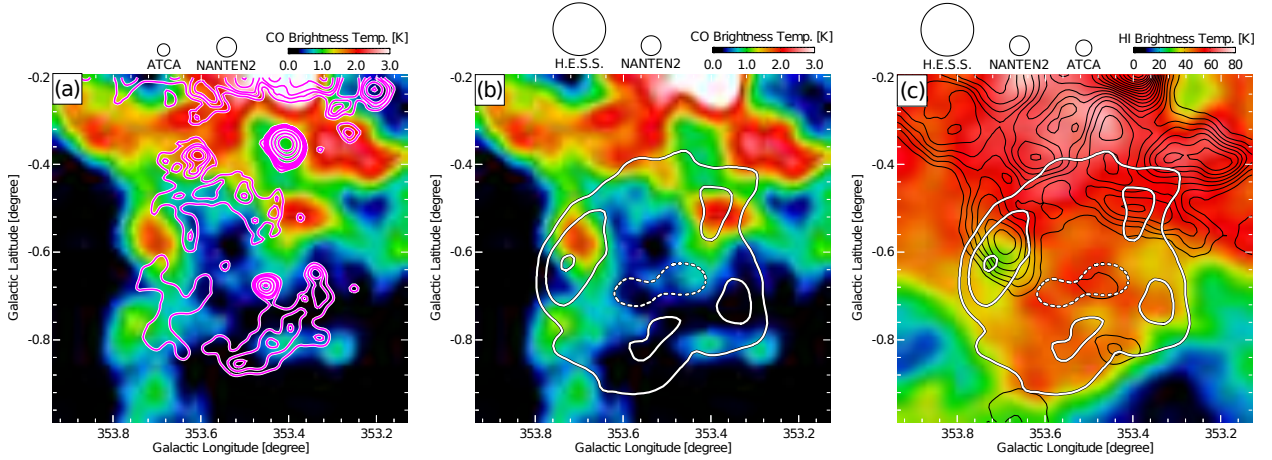}
\caption{(a)Averaged intensity map of ${}^{12}$CO($J$=1--0) in a velocity range from $-90\:\mathrm{km}\:\mathrm{s}^{-1}$ to $-75\:\mathrm{km}\:\mathrm{s}^{-1}$ is shown in color. Pink contours show 1.4 GHz continuum distribution and contour levels are 30, 45, 60, 100, 150, 200, 600, 1500, and 3000 $\mathrm{mJy}\:\mathrm{beam}^{-1}$. (b)Averaged intensity map of ${}^{12}$CO($J$=1--0) emission overlayed the TeV $\gamma$-rays contours. (c)Averaged intensity map of H{\sc i} in a velocity range from $-82\:\mathrm{km}\:\mathrm{s}^{-1}$ to $-80\:\mathrm{km}\:\mathrm{s}^{-1}$ emission overlaid on the TeV $\gamma$-rays and ${}^{12}$CO($J$=1--0). The minimum contour levels and intervals of ${}^{12}$CO($J$=1--0) are 0.6 K and 1.0 K, respectively.} 
\end{center}
\end{figure}

\begin{figure}
\begin{center}
\includegraphics[scale=0.4,angle=0,clip]{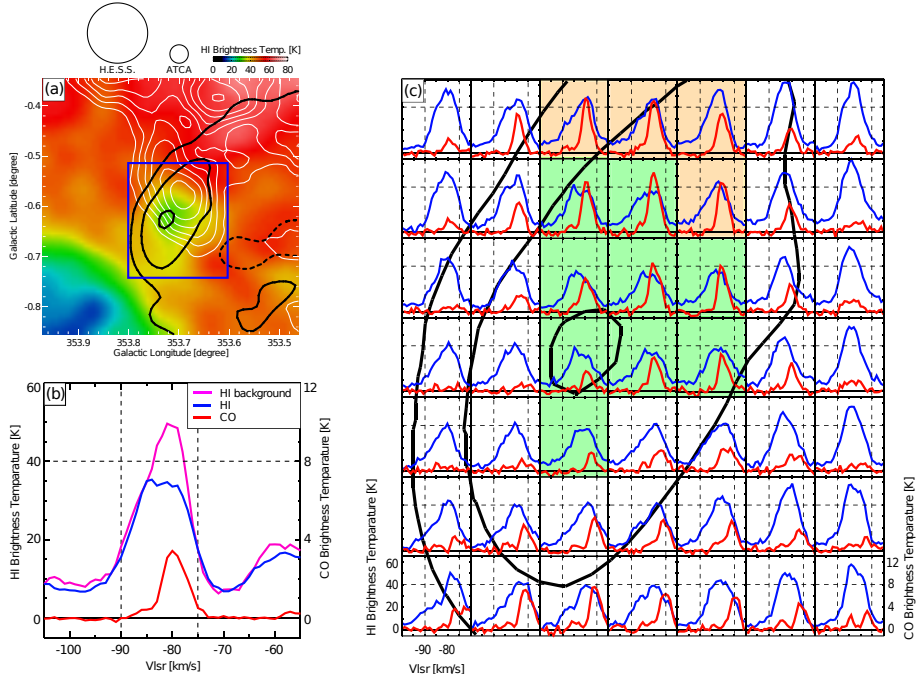}
\caption{(a)Averaged intensity map of H{\sc i} overlayed the TeV $\gamma$-rays and ${}^{12}$CO($J$=1--0) in a velocity range from $-82\:\mathrm{km}\:\mathrm{s}^{-1}$ to $-80\:\mathrm{km}\:\mathrm{s}^{-1}$ toward the northeast. Blue box is the area shown H{\sc i} profile and ${}^{12}$CO($J$=1--0) in Figure(c). (b)Averaged profile in absorption areas and no-absorption areas is shown in blue and pink, respectively. The red profile is ${}^{12}$CO($J$=1--0) in absorption areas. (c)Profile map of H{\sc i} and ${}^{12}$CO($J$=1--0) toward the northeast $\gamma$-rays peak is shown in blue and red, respectively. The northwest $\gamma$-rays is shown in black contours. The profile which H{\sc i} show the absorption is shown in green areas and the background H{\sc i} is shown in orange areas.} 
\end{center}
\end{figure}
\clearpage

\begin{figure}
\plotone{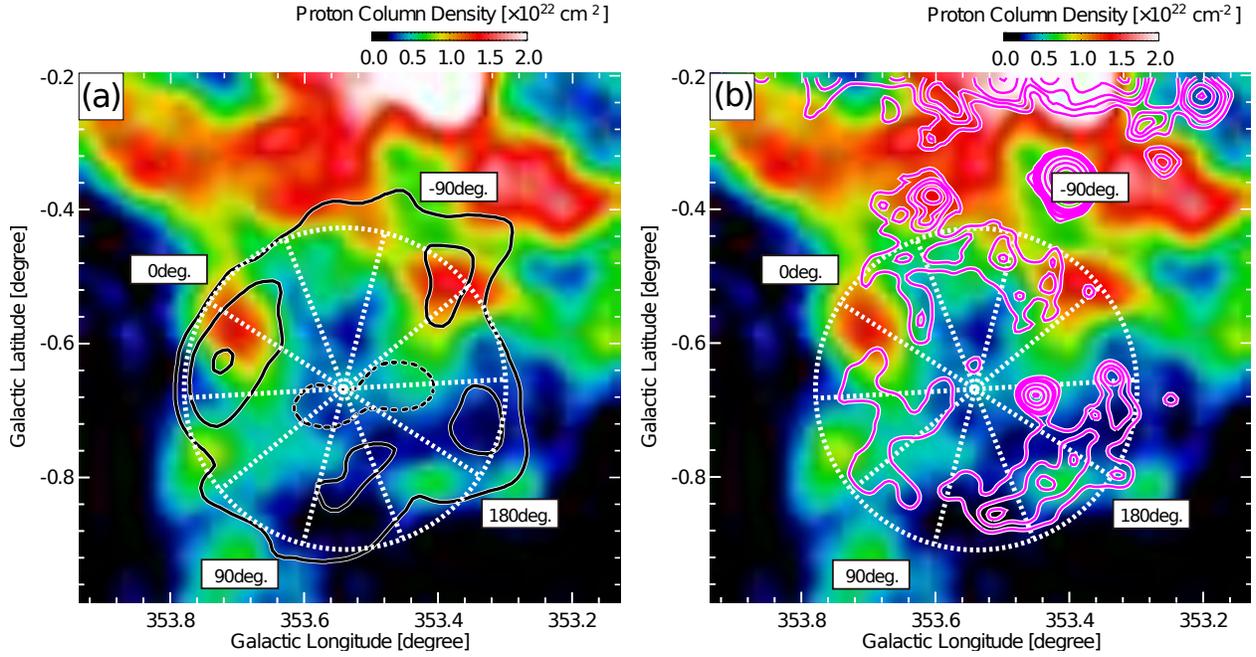}
\caption{Distributions of column density of the total ISM protons $N_\mathrm{p}(\mathrm{H_2}$+H{\sc i}) in a velocity range from $-90\:\mathrm{km}\:\mathrm{s}^{-1}$ to $-75\:\mathrm{km}\:\mathrm{s}^{-1}$. Contours are (a)the TeV $\gamma$-rays and (b) 1.4 GHz continuum distribution, respectively. The contours levels is the same as Figure 1(b) and Figure 5(a). The dashed white circle shows the position angle in azimuthal distribution (Figure 8(a)) and $0^\circ$ corresponds to North and $90^\circ$ to East in the equatorial coordinate.} 
\end{figure}
\clearpage

\begin{figure}
\plotone{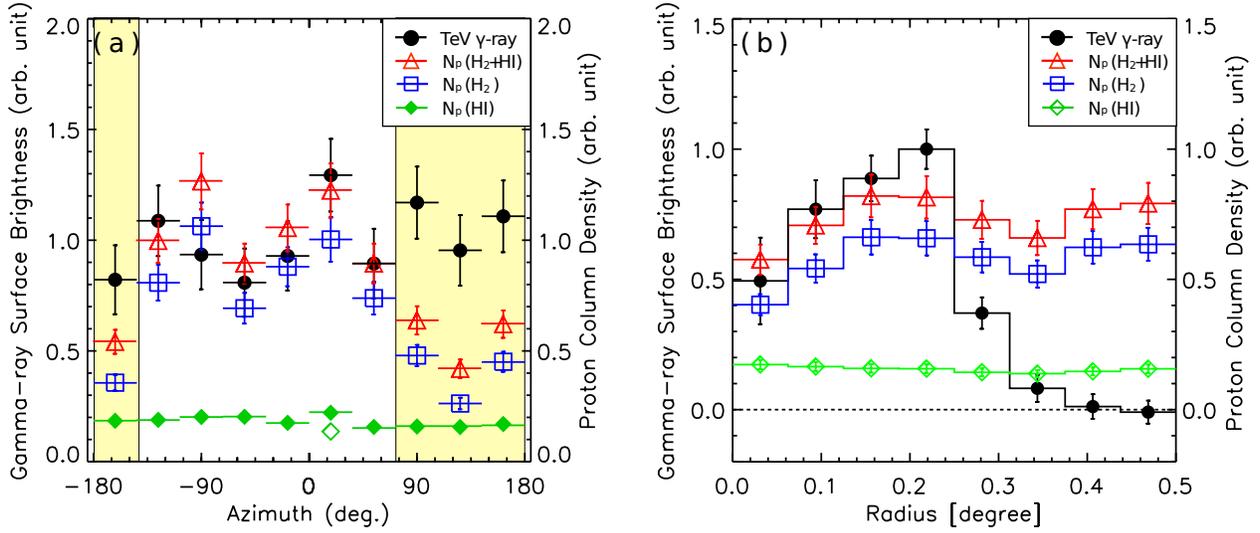}
\caption{(a) Azimuthal distribution of $N_\mathrm{p}(\mathrm{H_2})$, $N_\mathrm{p}$(H{\sc i}), $N_\mathrm{p}(\mathrm{H_2}$+H{\sc i}), and TeV $\gamma$-ray surface brightness. The proton column densities are averaged values between each area. The outlined green plot point at $18^\circ$ is $N_\mathrm{p}$(H{\sc i}) for the optically thin, and the solid green plot point is that for the optically thick for assumed $T_s$ = 33 K. (b)Radial distributions of averaged values of $N_\mathrm{p}(\mathrm{H_2})$, $N_\mathrm{p}$(H{\sc i}), $N_\mathrm{p}(\mathrm{H_2}$+H{\sc i}), and TeV $\gamma$-ray surface brightness.} 
\end{figure}
\clearpage

\begin{figure}
\begin{center}
\includegraphics[scale=1.2,angle=0,clip]{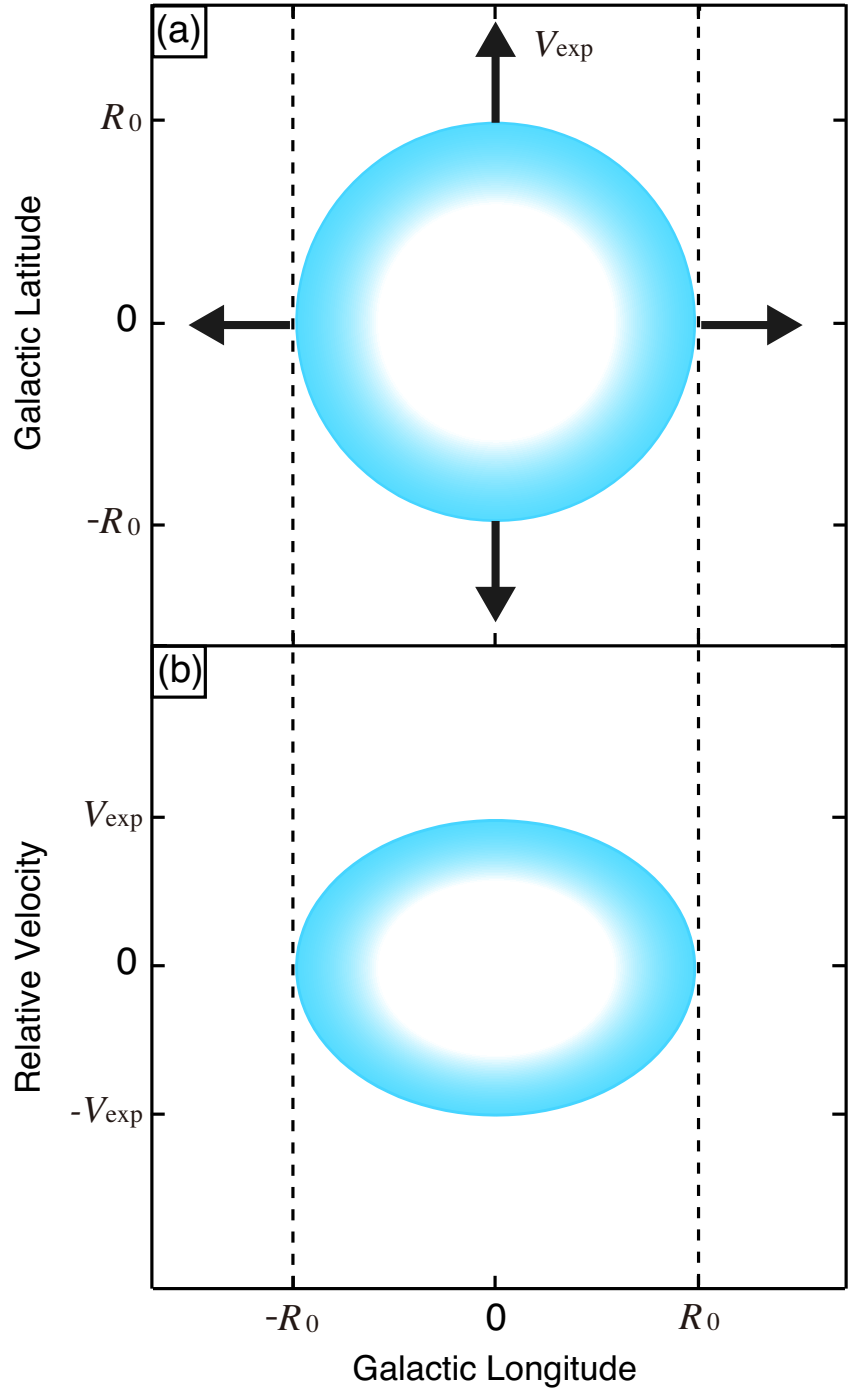}
\caption{(a) Schematic image of an isotropic expanding shell and (b) its position-velocity diagram.} 
\end{center}
\end{figure}


\begin{thebibliography}{}
\bibitem[Abramowski et al.(2011)]{2011A&A...531A..81H} Abramowski, A., Acero, F., et al.\ 2011, \aap, 531, A81

\bibitem[Abdo et al.(2011)]{2011ApJ...734...28A} Abdo, A.~A., Ackermann, M., Ajello, M., et al.\ 2011, \apj, 734, 28 

\bibitem[Acero et al.(2009)]{2009arXiv0907.0642A} Acero, F., P{\"u}hlhofer, G., Klochkov, D., et al.\ 2009, arXiv:0907.0642 

\bibitem[Aharonian et al.(2008)]{2008A&A...477..353A} Aharonian, F., Akhperjanian, A.~G., Barres de Almeida, U., et al.\ 2008, \aap, 477, 353 

\bibitem[Bamba et al.(2012)]{2012ApJ...756..149B} Bamba, A., P{\"u}hlhofer, G., Acero, F., et al.\ 2012, \apj, 756, 149 

\bibitem[Bell(1978)]{1978MNRAS.182..147B} Bell, A.~R.\ 1978, \mnras, 182, 147 

\bibitem[Bertsch et al.(1993)]{1993ApJ...416..587B} Bertsch, D.~L., Dame, T.~M., Fichtel, C.~E., et al.\ 1993, \apj, 416, 587 

\bibitem[Bissantz et al.(2003)]{2003MNRAS.340..949B} Bissantz, N., Englmaier, P., \& Gerhard, O.\ 2003, \mnras, 340, 949 

\bibitem[Blandford \& Ostriker(1978)]{1978ApJ...221L..29B} Blandford, R.~D., \& Ostriker, J.~P.\ 1978, \apjl, 221, L29

\bibitem[Cohen \& Davies(1976)]{1976MNRAS.175....1C} Cohen, R.~J., \& Davies, R.~D.\ 1976, \mnras, 175, 1 

\bibitem[Dame \& Thaddeus(2008)]{2008ApJ...683L.143D} Dame, T.~M., \& Thaddeus, P.\ 2008, \apjl, 683, L143

\bibitem[Dickey \& Lockman(1990)]{1990ARA&A..28..215D} Dickey, J.~M., \& Lockman, F.~J.\ 1990, \araa, 28, 215 

\bibitem[Ellison et al.(2010)]{2010ApJ...712..287E} Ellison, D.~C., Patnaude, D.~J., Slane, P., \& Raymond, J.\ 2010, \apj, 712, 287 

\bibitem[Englmaier \& Gerhard(1999)]{1999MNRAS.304..512E} Englmaier, P., \& Gerhard, O.\ 1999, \mnras, 304, 512 

\bibitem[Fukui et al.(2012)]{2012ApJ...746...82F} Fukui, Y., Sano, H., Sato, J., et al.\ 2012, \apj, 746, 82 

\bibitem[Fukui(2013)]{2013ASSP...34..249F} Fukui, Y.\ 2013, in Advances in Solid State Physics, Vol. 34, Cosmic Rays in Star-Forming
Environments, ed. D. F. Torres \& O. Reimer, 249

\bibitem[Fukui et al.(2014)]{2014arXiv1401.7398F} Fukui, Y., Okamoto, R., Yamamoto, H., et al.\ 2014, arXiv:1401.7398 

\bibitem[Fukui et al.(2014)]{2014arXiv1403.0999F} Fukui, Y., Torii, K., Onishi, T., et al.\ 2014, arXiv:1403.0999 

\bibitem[Giuliani et al.(2010)]{2010A&A...516L..11G} Giuliani, A., Tavani, M., Bulgarelli, A., et al.\ 2010, \aap, 516, L11 

\bibitem[Goldsmith et al.(2007)]{2007ApJ...654..273G} Goldsmith, P.~F., Li, D., \& Kr{\v c}o, M.\ 2007, \apj, 654, 273 

\bibitem[Green et al.(2009)]{2009ApJ...696L.156G} Green, J.~A., McClure-Griffiths, N.~M., Caswell, J.~L., et al.\ 2009, \apjl, 696, L156 

\bibitem[Green et al.(2011)]{2011ApJ...733...27G} Green, J.~A., Caswell, J.~L., McClure-Griffiths, N.~M., et al.\ 2011, \apj, 733, 27 

\bibitem[Haverkorn et al.(2006)]{2006ApJS..167..230H} Haverkorn, M., Gaensler, B.~M., McClure-Griffiths, N.~M., Dickey, J.~M., \& Green, A.~J.\ 2006, \apjs, 167, 230 

\bibitem[Inoue et al.(2012)]{2012ApJ...744...71I} Inoue, T., Yamazaki, R., Inutsuka, S.-i., \& Fukui, Y.\ 2012, \apj, 744, 71 

\bibitem[McClure-Griffiths et al.(2005)]{2005ApJS..158..178M} McClure-Griffiths, N.~M., Dickey, J.~M., Gaensler, B.~M., et al.\ 2005, \apjs, 158, 178 

\bibitem[Moriguchi et al.(2005)]{2005ApJ...631..947M} Moriguchi, Y., Tamura, K., Tawara, Y., et al.\ 2005, \apj, 631, 947 

\bibitem[Mulder \& Liem(1986)]{1986A&A...157..148M} Mulder, W.~A., \& Liem, B.~T.\ 1986, \aap, 157, 148 

\bibitem[Peters(1975)]{1975ApJ...195..617P} Peters, W.~L., III 1975, \apj, 195, 617 

\bibitem[Ridge et al.(2006)]{2006AJ....131.2921R} Ridge, N.~A., Di Francesco, J., Kirk, H., et al.\ 2006, \aj, 131, 2921 

\bibitem[Rodriguez-Fernandez \& Combes(2008)]{2008A&A...489..115R} Rodriguez-Fernandez, N.~J., \& Combes, F.\ 2008, \aap, 489, 115 

\bibitem[Sanna et al.(2009)]{2009ApJ...706..464S} Sanna, A., Reid, M.~J., Moscadelli, L., et al.\ 2009, \apj, 706, 464 

\bibitem[Sano et al.(2010)]{2010ApJ...724...59S} Sano, H., Sato, J., Horachi, H., et al.\ 2010, \apj, 724, 59 

\bibitem[Sevenster(1999)]{1999MNRAS.310..629S} Sevenster, M.~N.\ 1999, \mnras, 310, 629 

\bibitem[Tanaka et al.(2008)]{2008ApJ...685..988T} Tanaka, T., Uchiyama, Y., Aharonian, F.~A., et al.\ 2008, \apj, 685, 988 

\bibitem[Tian et al.(2008)]{2008ApJ...679L..85T} Tian, W.~W., Leahy, D.~A., Haverkorn, M., \& Jiang, B.\ 2008, \apjl, 679, L85 

\bibitem[Tian et al.(2010)]{2010ApJ...712..790T} Tian, W.~W., Li, Z., Leahy, D.~A., et al.\ 2010, \apj, 712, 790 

\bibitem[Uchiyama et al.(2007)]{2007Natur.449..576U} Uchiyama, Y., Aharonian, F.~A., Tanaka, T., Takahashi, T., \& Maeda, Y.\ 2007, \nat, 449, 576 

\bibitem[van Woerden et al.(1957)]{1957CRAS..244.1691V} van Woerden, H., Rougoor, G.~W., \& Oort, J.~H.\ 1957, Academie des Sciences Paris Comptes Rendus, 244, 1691 

\bibitem[Yoshiike et al.(2013)]{2013ApJ...768..179Y} Yoshiike, S., Fukuda, T., Sano, H., et al.\ 2013, \apj, 768, 179 

\bibitem[Zirakashvili \& Aharonian(2010)]{2010ApJ...708..965Z} Zirakashvili, V.~N., \& Aharonian, F.~A.\ 2010, \apj, 708, 965 


\end{thebibliography}
\end{document}